    \documentclass{emulateapj}
    
    \usepackage{natbib}
    \usepackage{graphicx,subfigure} 
    \usepackage{amsmath}

    \newcommand\lsim{\mathrel{\rlap{\lower4pt\hbox{\hskip1pt$\sim$}}
            \raise1pt\hbox{$<$}}}
    \newcommand\gsim{\mathrel{\rlap{\lower4pt\hbox{\hskip1pt$\sim$}}
            \raise1pt\hbox{$>$}}}
    
    \newcommand \mum{ \mu{\rm m} }
    \newcommand \charsig{ \overset{\sim}{\sigma} }
    \newcommand \Msun{ {\rm M}_\odot }

    \newcommand \OmegaIGM{ \Omega_{d}^{\rm IGM} }

\begin{document}
    
    \title{X-ray Scattering Echoes and Ghost Halos from the Intergalactic Medium:\\ Relation to the nature of AGN variability}
    \author{Lia Corrales}
    \affil{MIT Kavli Institute for Astrophysics and Space Research,}
    \affil{Columbia University}

    \begin{abstract}
        
    X-ray bright quasars might be used to trace dust in the circumgalactic and intergalactic medium through the phenomenon of X-ray scattering, which is observed around Galactic objects whose light passes through a sufficient column of interstellar gas and dust.  Of particular interest is the abundance of grey dust larger than $0.1~\mum$, which is difficult to detect at other wavelengths.  To calculate X-ray scattering from large grains, one must abandon the traditional Rayleigh-Gans approximation.  The Mie solution for the X-ray scattering optical depth of the Universe is $\sim 1\%$.  This presents a great difficulty for distinguishing dust scattered photons from the point source image of {\sl Chandra}, which is currently unsurpassed in imaging resolution.  The variable nature of AGN offers a solution to this problem, as scattered light takes a longer path and thus experiences a time delay with respect to non-scattered light.  If an AGN dims significantly ($\gsim 3$~dex) due to a major feedback event, the {\sl Chandra} point source image will be suppressed relative to the scattering halo, and an X-ray echo or ghost halo may become visible.  I estimate the total number of scattering echoes visible by {\sl Chandra} over the entire sky: $N_{\rm ech} \sim 10^3 (\nu_{\rm fb}/{\rm yr}^{-1})$, where $\nu_{\rm fb}$ is the characteristic frequency of feedback events capable of dimming an AGN quickly.  
        
    \end{abstract}

    \section{Introduction}
    \label{sec:Introduction}

Several lines of evidence suggest that galaxies do not evolve in closed boxes; a noticeable fraction of material can escape galaxy disks or bulges where the vast majority of luminous baryonic matter resides.  Metals are the dominant tracers of star-processed material, a fraction of which is locked up in dust.  
Dust may also play an essential role in galaxy feedback by giving the interstellar medium (ISM) the opacity necessary to launch radiation pressure driven winds \citep{MQT2005}.  
Assessing the abundance of extragalactic dust will thereby provide insight for the evolution of the Universe, from the production of metals by star formation, to the mechanisms responsible for the structure of galaxies.
The purpose of this work is to evaluate whether the phenomenon of X-ray scattering can be used to trace dust in the circumgalactic (CGM) and intergalactic medium (IGM).

    Distant X-ray bright AGN proffer the opportunity to probe dust in the IGM or intervening dust reservoirs \citep{Evans1985, ME1999, Corrales2012}.  
    Dust grains scatter X-ray light over arcminute scales, producing a diffuse halo image around bright point sources.  The first X-ray scattering halo was imaged by the Einstein Observatory around GX 339-4 \citep{Rolf1983}.  Scattering halos are now ubiquitously found around objects whose light passes through a significant ISM column, including Galactic X-ray binaries \citep[e.g.][]{Smith2008, Valencic2009, Tiengo2010, Xiang2011} and extragalactic sources whose sight lines happen to lie behind dense regions of Milky Way ISM \citep{Vaughan2006}.

    Historically, X-ray scattering has been treated using the Rayleigh-Gans approximation (RG hereafter), which is valid when electromagnetic waves suffer negligible phase shift while traversing a dust grain \citep{vdHbook, Overbeck1965, Hayakawa1970, MG1986}.  From the perspective of high energy light, the condensed matter in dust grains is roughly a ball of electrons, so the Drude approximation to the complex index of refraction can also be applied \citep[][hereafter SD98]{SD1998}.  
    The RG-Drude approximation yields a cross-section that is exceptionally sensitive to the size of the dust grains ($a$) and increasingly efficient for softer X-rays, $\sigma_{\rm RG} \propto a^4 E^{-2}$ \citep[e.g.][SD98]{MG1986}.

    RG-Drude's sensitivity towards grain size makes it particularly tempting to use in searching for large intergalactic dust, for which there exists both theoretical and observational motivation.  Balancing the forces of radiation pressure, gravity, and gas drag, several works show that mainly dust grains $\gsim 0.1~\mum$ can reach distances a few to 100s of kpc from a galaxy disk {\sl and} survive entrainment through hot halo gas \citep{Barsella1989, Ferrara1990, Ferrara1991, Davies1998}.  
    Such grains are `grey',  blocking UV and optical wavelengths uniformly, and altering the perceived brightness of extragalactic objects without leaving the tell-tale sign of reddening.  
    %
    Grey grains leaking into the IGM will cause systematic shifts for photometric measurements of Type Ia SNe in dark energy surveys, altering their figure of merit \citep[e.g.~][]{Virey2007, Suzuki2012}.  A magnitude shift $\delta m$ on the order of a percent is large enough to shift the determination of cosmological parameters by $\delta w \sim \delta m$ and $\delta \Omega_M \sim -\delta m$, which is on the level of the precision sought after by the next generation of cosmological surveys \citep{ZhangP2008, Menard2010b, Corasaniti2006}.

    \citet{Petric2006} made the first attempt to image an IGM dust scattering halo using a {\sl Chandra} observation of the $z=4$ object QSO 1508+5714.  From the absence of a detectable halo, they concluded that the density of IGM dust must be $\OmegaIGM \leq 2 \times 10^{-6}$, relative to the critical density.  
    However, as one looks towards larger dust grains and softer X-ray energies where the RG-Drude cross-section increases dramatically, the approximation begins to break down.  SD98 showed that this occurs roughly when $a_{\mum} \leq E_{\rm keV}$ is no longer true.  Cases that violate this rule-of-thumb need to use the Mie scattering solution, which can be calculated numerically \citep[e.g.][]{BHbook, Wiscombe1980}.  
    Because the RG-Drude approximation overestimated the scattering opacity from grey dust by at least a factor of 10, \citet[][hereafter CP12]{Corrales2012} showed that the limit on the abundance of intergalactic dust found by \citet{Petric2006} could be relaxed to $\OmegaIGM \lsim 10^{-5}$.

    This paper follows up on the work of CP12 by evaluating scattering in the Mie regime and comparing scattering halo surface brightness profiles to the {\sl Chandra} point spread function, or PSF (Section~\ref{sec:MieScatteringHalos}).  The dust scattering optical depth of the universe is $\sim 1\%$, which is comparable to the brightness of the Chandra PSF wings (outside of $1-2''$).  The only way to view an IGM dust scattering halo is if the point source brightness is suppressed in some way.  Since scattered light is time delayed relative to the point source, AGN that suddenly decrease in brightness may leave behind a scattering echo.  I will evaluate the quasar population and comment on the observability of X-ray scattering echoes with respect to the {\sl Chandra} background (Section~\ref{sec:ScatteringEchoes}).  A summary of the results and review of implications, should a scattering echo be observed, is presented in Section~\ref{sec:Conclusion}.

    \section{Intergalactic Scattering in the Mie Regime}
    \label{sec:MieScatteringHalos}

From Equation~19 of CP12, the optical depth of the Universe to X-ray scattering through a uniformly distributed medium is:
\begin{equation}
\label{eq:TauUniverse}
	\tau_x (z_s, E) = \int \int_0^{z_s} \sigma_{E,z}\ n_d\ (1+z)^2\ \frac{c\ dz}{H}\ da
\end{equation}
where $z_s$ is the redshift of the central X-ray point source, $E$ is the observed photon energy, $\sigma_{E,z}$ is the scattering cross-section of a photon with energy $E (1+z)$, $H = H_0 \sqrt{ \Omega_\Lambda + \Omega_m (1+z)^3 }$, and $n_d$ is the comoving grain size distribution, presumed constant such that the true density at any given redshift is $n_d (1+z)^3$.  All calculations in this work use standard $\Lambda$CDM cosmology with $\Omega_m = 0.3$, $\Omega_\Lambda = 0.7$, $H_0 = 75$~km/s/Mpc, and zero curvature.  

%
I derive the number density as a function of grain size using the constraint:
\begin{equation}
	\int \frac{4}{3} \pi a^3\ \rho_g\ n_d\ da = \OmegaIGM\ \rho_c
\end{equation}
where $\rho_g$ is the grain material density and $\rho_c$ is the cosmological critical density.  Typical values for $\rho_g$ are 2.2~g~cm$^{-3}$ and 3.8~g~cm$^{-3}$ for graphite and silicate grains, respectively \citep{DraineBook}.  
    %
    This work also draws upon the canonical result of \citet[][hereafter MRN]{MRN1977}, who found that a mix of graphite and silicate grains following a power law size distribution can describe the extinction curves of local interstellar dust.  
    Their fiducial power law, $n_d \propto a^{-3.5}$, is used throughout this work, but the end points of the distribution are varied to approximate an extinction curve that is more red ($0.005~\mum \leq a \leq 0.3~\mum$) or more grey ($0.1~\mum \leq a \leq 1~\mum$).  I will investigate grain populations that are made up of purely silicate or purely graphite grains in order to bracket the possible solutions that come from a mixture of the two.  
    In calculations invoking a single grain size, the delta function $n_d \propto \delta (a - a_g)$ will be used, where $a_g$ is the grain radius of interest.
    
\begin{figure*}[tp]
    	\centering
    	\includegraphics[scale=0.71, trim=65 0 0 10]{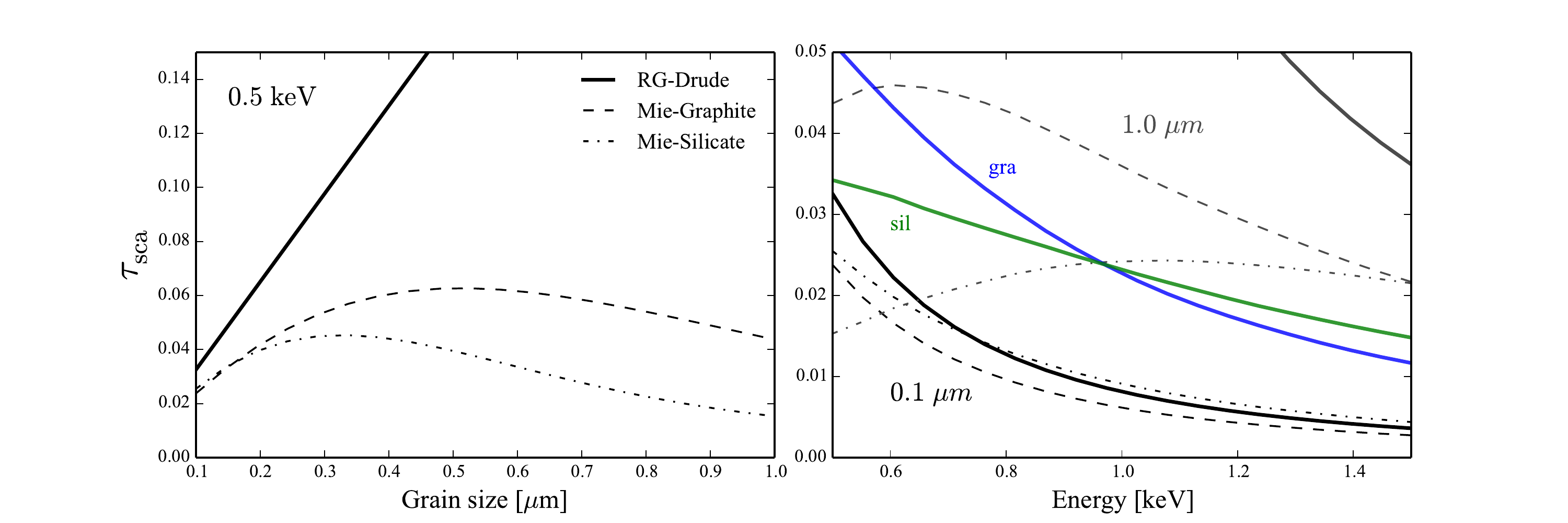}
    	\caption{The optical depth to X-ray scattering is plotted for a $z = 2$ point source and $\OmegaIGM = 10^{-5}$.  
    	{\sl Left:} 
    	The 0.5~keV Mie scattering solution for a population of singly sized grains is shown using graphite ({\sl dashed}) and silicate ({\sl dot-dash}) materials.  
    	{\sl Right:} The scattering optical depth for $0.1\ \mum$ ({\sl dark curves}) and $1.0\ \mum$ ({\sl light curves}) as a function of energy.  
	The Mie solution for a grain size distribution with $0.1~\mum \leq a \leq 1~\mum$ is shown for silicate ({\sl green curve}) and graphite ({\sl blue curve}). }
    	\label{fig:CosmoTauxMie}
    \end{figure*}

    Using the upper limit $\OmegaIGM = 10^{-5}$ and a point source at $z_s = 2$, Figure~\ref{fig:CosmoTauxMie} shows the Mie scattering optical depth of IGM dust from silicate and graphite.  Dielectric constants for both grain types were taken from \citet{Draine2003b} and used in conjunction with the publicly available \citet{BHbook} Mie scattering code.  This work confirms the conjecture from CP12 that the RG-Drude approximation is a factor of 3-10 too large when investigating soft X-ray scattering from grey dust, but is still roughly valid if the dust has a characteristic size around $0.1~\mum$.  Figure~\ref{fig:CosmoTauxMie} also shows the Mie scattering solution for a grey distribution of graphite and silicate dust grains, which will be used later to calculate the scattering halo intensity from intergalactic dust.

    Figure~\ref{fig:CosmoTauxMie} demonstrates that, for the most optimistic conditions, only 1-5\% of $z=2$ quasar light will scatter through the diffuse intergalactic medium.  This is roughly the same fraction of photons that will scatter off micro-roughness in the {\sl Chandra} mirrors, which provide the highest imaging resolution currently available in X-ray instrumentation.  The ability to distinguish IGM or CGM scattered light from the {\sl Chandra} PSF and background depends on a number of factors, the subject of the remainder of this work.  To begin the investigation, one must calculate the intensity profile of scattered light.
    
    
The integral for the scattering halo intensity ($I_h$) is directly proportional to the apparent flux ($F_a$) of the central point source.  Following CP12, I define the normalized intensity $d\Psi/d\Omega \equiv I_h / F_{a}$, yielding
    \begin{equation}
    \label{eq:HaloIntensity}
    	\frac{d \Psi}{d \Omega} (\alpha, E) = 
    	\int \int_0^{z_s} N_d\ \xi(z)\  
    	\frac{(1+z)^2}{x^2}\ 
    	\left[ \frac{d\sigma_{E,z}} {d\Omega} \right]_{\alpha/x}
    	\frac{c\ dz}{H}\ da
    \end{equation}
    where $\alpha$ is the observation angle with respect to the point source center, $z_s$ is the redshift of the source, $N_d$ is the grain size distribution in terms of column density\footnote{
    As before it is presumed that $N_d \propto a^{-3.5}$, except for examination of a single grain size, in which case $N_d \propto \delta (a - a_g)$.
    } with the dimensionless function $\xi(z)$ parameterizing the dust spatial distribution along the line of sight, $d\sigma_{E,z}/d\Omega$ is the differential cross-section evaluated at scattering angle $\alpha/x$ using the energy relevant to the dust grain at $z$, and $x$ is the dimensionless distance
    \begin{equation}
    \label{eq:DefineX}
    	x = \frac{ \int_z^{z_s} c\ dz / H }{ \int_0^{z_s} c\ dz / H }.
    \end{equation}
    The parameter $\xi=1$ in the case of uniformly distributed dust with constant co-moving density and $\xi = \delta(z - z_g)$ for a dust screen at redshift $z_g$.  For a full derivation of dust scattering in a cosmological context, see CP12. 

    \subsection{Scattering halo intensity from IGM dust}
    \label{sec:IGMhalos}

    The amount of dust and metals predicted from the star formation history of the Universe does not currently match that observed thus far.  
    Dubbed the `missing metals problem', approximately 65\% of metals in the $z \sim 2$ Universe can be accounted for from observations of galaxies and intergalactic absorption systems \citep{Bouche2007}.  
    %
    %
    There is also a deficit between the predicted and observed amount of dust found in galaxy disks and diffuse regions of galaxy halos combined \citep[][hereafter MF12]{MF2012}.  
    %
    They predict an excess of dust $\OmegaIGM \approx 3 \times 10^{-6}$ at $z \approx 0.3$, the epoch that contributes most to an intergalactic scattering halo (CP12). 
    
    \citet{Shull2014} estimate that the low redshift IGM contains roughly 10\% of all cosmic metals, implying an IGM metal density $\Omega_Z^{\rm IGM} \approx 10^{-5}$ in terms of the critical density.  This value is in agreement with the results of MF12 if the depletion factors are 30\%, which is common.  In fact, recent cosmological simulations that include the effect of radiation pressure driven galactic winds are able to reproduce observations of CGM dust \citep[][hereafter MSFR10]{Menard2010} using a dust-to-metals ratio of 0.29 \citep{Zu2011}.  
   From here forward I take $\OmegaIGM = 3 \times 10^{-6}$ as a fiducial mass density for deriving the number density of dust grains $N_d$ used in Equation~\ref{eq:HaloIntensity}.

    Figure~\ref{fig:MieScatteringHalos} shows the normalized scattering halo intensity using $z_s = 2$ and the Mie scattering solution for $d\sigma_{E,z}/d\Omega$ in Equation~\ref{eq:HaloIntensity}.  I found very little difference between graphite and silicate scattering halos, so only graphite material is shown for scattering off uniformly distributed IGM dust.
    In the Mie regime, the singly-sized $1~\mum$ dust population does not produce a significantly brighter scattering halo than that caused by a grey dust grain size distribution.

    \subsection{Scattering halo intensity from CGM dust}
    \label{sec:CGMhalos}

    CP12 evaluated various sources and reservoirs of intergalactic dust, including absorption line systems.  This work considers one additional reservoir: the circumgalactic medium, which I will use to describe any dusty material residing within 300~kpc of a galaxy disk.  
    
    There is plenty of direct observational evidence that dust escapes galaxy disks to pollute the CGM.  
    %
    %
    Dust is found out to 9 kpc around starburst galaxy M82, imaged via polarized scattering \citep{Schmidt1976}, polycyclic aromatic hydrocarbon (PAH) emission from a warm gas component mixed with the superwind \citep{Engelbracht2006}, and far-infrared emission from a cold gas component mixed in with tidally stripped material \citep{Roussel2010}.  
    \citet[][hereafter HK14]{HodgesKluck2014} also observe large angle UV scattering from dust 10-20~kpc above the mid-plane of 20 early type galaxies.

    Using over $10^7$ objects from the Sloan Digital Sky Survey, MSFR10 calculate the average quasar color as a function of impact parameter to a foreground galaxy.  They find a reddening trend out to distances as large as 1 Mpc from galaxy centers, leading them to conclude that the mass of dust in galactic halos is comparable to that in galaxy disks.  Recent work by \citet{PMC2014} repeats the experiment with galaxy-galaxy pairs and comes to a similar mass estimate.  
    
    MSFR10 calculate the average CGM extinction as a function of impact parameter to a nearby foreground galaxy.  
    Inverting Equation~30 of their paper to solve for the total dust mass column yields:
    \begin{equation}
    \label{eq:GalHaloDustMass}
    	M_d \approx 4 \times 10^{-3} \ 
    	\kappa_{\rm V}^{-1} \ 
    	\left( \frac{r_p}{100\ h^{-1} \ {\rm kpc}} \right)^{-0.86}
    \end{equation}
    where $r_p$ is the impact parameter to a foreground galaxy, $\kappa_{\rm V}$ is the V-band extinction opacity, and $h$ is the Hubble constant in units of 100 km/s/Mpc.
    I approximate the scattering from dusty CGM with an infinitely large, homogeneous screen at $z_g = 0.5$ containing a dust mass column equivalent to an $r_p \sim 100$~kpc sight line.  Due to the spatial distances involved, 100~kpc is associated with observation angles $\sim 10''$, which is about the visible extent of intergalactic scattering halos (CP12).

    MSFR10 also found that the slope of the quasar colors in their study were roughly consistent with local ISM.  However, they prefer extinction curves similar to the Small Magellanic Cloud (SMC) due to the fact that Mg II absorbers, which they believe are tracers of galaxy outflow, show reddening curves that lack the 2175\AA\ feature often attributed to PAHs and small graphite grains \citep{York2006}.  SMC extinction curves also lack the 2175\AA\ bump.  HK14 also found that the spectral energy distribution of UV reflection from CGM dust fits well with SMC-like dust in many cases.

\begin{figure}[tp]
    	\centering
    	\includegraphics[scale=0.48, trim=20 20 0 30]{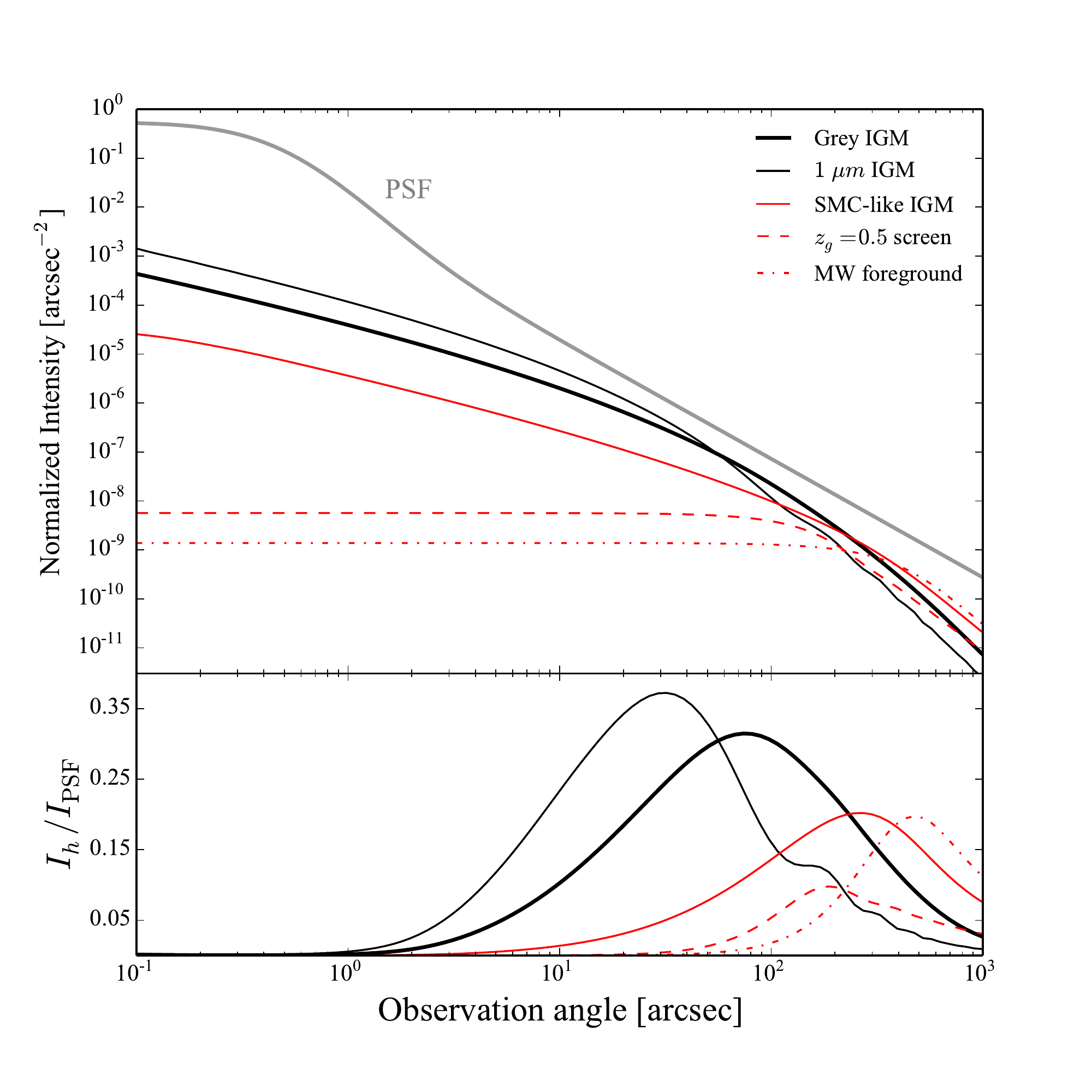}
    	\caption{ 
	({\sl Top:})  The 1~keV intensity for intergalactic scattering halos around a $z_s=2$ source in comparison to the {\sl Chandra} PSF template ({\sl top}), and their ratios ({\sl bottom}).  
    	The Mie scattering halo from uniformly distributed IGM dust ($\OmegaIGM = 3 \times 10^{-6}$) is shown for a grey $0.1-1\ \mum$ grain size distribution ({\sl thick black curve}), which is not significantly different from the distribution from singly-sized $1\ \mum$ grains ({\sl thin black curve}).  If IGM dust followed a more typical SMC-like dust distribution that cuts off at $0.3\ \mum$, the scattering profile brightness would be roughly an order of magnitude dimmer ({\sl solid red curve}).  
    	A dust column consistent with the results of M10 ($M_d = 3 \times 10^{-7}$ g cm$^{-2}$) around a $z_g=0.5$ foreground galaxy is simulated with a screen containing SMC-like dust ({\sl red dashed curve}).  
    	The scattering halo associated with a hypothesized mass of foreground Galactic dust  ($N_H = 10^{19}$ cm$^{-2}$), presumably from the Milky Way CGM, is also plotted using an SMC-like grain size distribution ({\sl red dashed-dotted curve}). 
	({\sl Bottom:}) The halo-to-PSF brightness ratio for the situations described above.  To obtain a robust detection of intergalactic dust, the PSF must be suppressed by a significant factor.
	}
    	\label{fig:MieScatteringHalos}
    \end{figure}

    Following their lead, I use the simplest approximation that produces an SMC-like extinction curve -- an MRN distribution of silicate dust \citep{Pei1992} with sizes $0.005~\mum \leq a \leq 0.3~\mum$  -- to model the scattering halo contribution from a foreground galaxy (Figure~\ref{fig:MieScatteringHalos}).  
    Scattering off CGM dust that may be occupying the halo of our own Galaxy is also shown, using a screen position of $z_g = 0$ and the dust mass equivalent for $N_{\rm H} = 10^{19}$~cm$^{-2}$.  Both cases are dramatically dimmer than the scattering one might expect from uniformly distributed, early-enriched IGM, demonstrating that at least the contribution from dust in the Milky Way halo should be insignificant.

On the other hand, the dust distribution from a foreground galaxy is not an azimuthally symmetric, homogenous scattering screen.  For quasar sight lines lying within 1~Mpc of a foreground galaxy, MSFR10 observe a color excess that spans two orders of magnitude.  Since the scattering halo intensity is directly proportional to dust abundance, one might expect significant variation in the amplitude of the dashed red curve shown in Figure~\ref{fig:MieScatteringHalos}, depending on the impact parameter of the quasar sight line.  A more detailed study of X-ray scattering from the CGM of a foreground galaxy requires examination of a non-homogeneous screen for a variety of redshifts and impact parameters, including the additional absorption that scattered light incurs in comparison to non-scattered light.  This is beyond the scope of this work.  I wish only to demonstrate that X-ray scattering is more suited to probe a near-homogeneous, diffuse component of the IGM.   Except for very special quasar-galaxy pairs, scattering from foreground galaxy dust should be less of a concern.

    The SMC-like dust found thus far by MSFR10 and HK14 produces a steep reddening signature, which is 
contrary to the hypothesis entertained by CP12 -- that dust escaping galaxy disks needs to be grey ($\gsim 0.1~\mum$) to receive sufficient radiation pressure and survive hot gas.  One can offer a variety of explanations for the CGM dust populations found thus far: (i) dust removed from the galaxy disk via feedback mechanisms may be subject to grain shattering by shocks, producing a distribution that is similar to that of the ISM but lacking in large dust grains \citep{JTH1996}; or (ii) small grains are trapped in galaxy halos  due to an equilibrium between gravity and radiation pressure while grains $\gsim 0.1~\mum$ escape to greater distances \citep{Greenberg1987,Ferrara1990,Barsella1989,Davies1998}.  Note that, once they get there, grains at relative rest with the low-density 
    and hot 
    environment of the CGM can survive for a Hubble time, $\sim 10^{10} (a/0.1\ \mum)$~years \citep{DS1979, DraineBook}.

    A final point rests in the fact that observational bias of the techniques used thus far may simply prohibit our ability to constrain the abundance of grey CGM dust.  Color studies relying on optical to UV wavelengths are relatively blind to large dust grains, and scattering cross sections are strongest for grains with a size comparable to the wavelength of interest \citep[e.g.][]{coreshine}.

    Due to a lack of sufficient information, I will refrain from speculating on the mass of grey dust within a few 100~kpc of galaxy disks.  However, if the red dust found thus far in the CGM is indicative of that leaking into the IGM, Figure~\ref{fig:MieScatteringHalos} shows that the scattering halo from SMC-like IGM dust is a factor of 10 dimmer than that produced by grey dust.

    \subsection{Comparison to the {\sl Chandra} PSF}
    \label{sec:ChandraPSF}

    {\sl Chandra} provides imaging resolution unsurpassed by current X-ray telescopes, confining 95\% of point source light to a 2 pixel (approximately $1''$) region.\footnote{{\sl Chandra}~Proposers'~Observatory~Guide (CPOG) http://cxc.harvard.edu/proposer/POG/}  
    This resolution is ideal for studying X-ray scattering halos from IGM dust, which have half-light radii $\sim 15-30''$ (CP12).  However Figure~\ref{fig:CosmoTauxMie} shows that we might expect only $\sim 1\%$ of soft X-ray light to scatter from IGM dust, making essential a careful understanding of the PSF wings -- caused by scattering off micro-abrasions on the instrument mirror -- which can easily be confused with dust scattered photons.  
    To create a template for the {\sl Chandra} PSF wings, I chose the bright Galactic X-ray binary Her X-1, which is used frequently for PSF calibration \citep{Gaetz2004,Gaetz2010}.  Her X-1 has a very low ISM column \citep[$N_{\rm H} \approx 8 \times 10^{19}$~cm$^{-2}$, using][]{SFD1998, Bohlin1978}, implying $\tau_{\rm sca}(1\ {\rm keV}) \approx 0.5\%$ from Galactic dust.

    I fit the 0.8-1.2~keV zeroth order image from a {\sl Chandra} HETG observation of Her X-1 (ObsId 2749) with a model containing a beta profile for the PSF core, power law wings (modified with an exponential decay term at small radii to create a smooth core-wing transition), and a constant background.  Figure~\ref{fig:MieScatteringHalos} plots the background-free PSF model in grey.  
    Even though the level of Galactic scattering expected for Her X-1 is on the order of intergalactic scattering for quasars, the background of ObsId 2749 far outshines the Her X-1 scattering halo and dominates the surface brightness profile at radii $> 50''$.  
    \citet{Gaetz2010} use deeper grating-free observations of Her X-1 to show that the PSF wings curve slightly at much larger radii, but this detail is not significant for the purposes of this work.  Section~\ref{sec:ScatteringEchoes} will show that the observable size of an IGM scattering halos is typically $10-20''$, negating the need to model the PSF wings any further.

    In all of the cases evaluated so far, the {\sl Chandra} PSF outshines scattering from intergalactic dust.  
    The bottom portion of Figure~\ref{fig:MieScatteringHalos} demonstrates that the scattering halo surface brightness will be, at most, 20-30\% of the PSF brightness.  However, the peak in $I_h/I_{\rm PSF}$ occurs in the area where the instrument background is more likely to dominate (around $100''$ for bright sources).
    The extremely low scattering optical depths also require unprecedented knowledge of the PSF wings, which is affected by a number of other systematic uncertainties, such as the telescope effective area and pileup.  It is therefore highly unlikely that, in the near future, the {\sl Chandra} PSF will be accurately calibrated to within 1\% of the point source brightness.  
    
    It may only be possible to view a scattering halo from IGM dust if the PSF is suppressed in some way.  
    Fortunately, the accretion processes powering AGN are known to be highly variable.  Since scattered light takes a longer path, the flux from the scattering halo is delayed by a characteristic timescale.  Incoming scattered light is proportional to the average flux of the point source over a period of time $\delta t$ in the past.  
    We might therefore see an X-ray scattering echo from IGM dust if an AGN has suddenly dropped in luminosity, i.e.~$F_{ps}(t) << F_{ps}(t - \delta t)$.      An AGN that dims suddenly may also fall below an instrumental detection threshold, leaving behind a ring of low surface brightness -- a ghost halo -- with no point source at the center.

Judging from the bottom portion of Figure~\ref{fig:MieScatteringHalos}, the point source image might have to be reduced by a factor of 20 to bring the scattering halo surface brightness levels on par with the PSF wings.  To be more conservative, I choose a factor of 1000 decrease in AGN luminosity as a benchmark.  In this case, scattering that would have been $\sim 1\%$ of the PSF brightness (e.g.~scattering from SMC-like IGM dust at $10''$) will instead be 10 times brighter than the point source image, allowing for a $3\sigma$ detection.  This situation would allow for unambiguous confirmation (or ruling out) of all three of the IGM dust cases examined above.
    The next section will evaluate the flux and variability requirements necessary for an AGN to produce X-ray scattering echoes brighter than the typical {\sl Chandra} background.

    \section{X-ray Scattering Echoes and ``Ghost'' Halos}
    \label{sec:ScatteringEchoes}
    
    Scattered light observed at angle $\alpha$ has an associated time delay, depending on the distance to the scattering material:
    \begin{equation}
    \label{eq:DeltaTgeneral}
    	\delta t = \frac{D_R}{c}\ \alpha^2 (1 - x)
    \end{equation}
    where $D_R$ is the radial distance to the source and $x$ is the position of the scatterer along the sightline as defined in Equation~\ref{eq:DefineX} \citep[CP12, see also][]{TS1973,ME1999}.  
    \citet{MG1986} show that the angular dependence of the RG differential scattering cross section can be approximated with a Gaussian of characteristic width 
    \begin{equation}
    \label{eq:Charsig}
    	\charsig = 10.4' \ \left( \frac{a}{0.1\ \mum}\right)^{-1} \ \left( \frac{E}{{\rm keV}} \right)^{-1} .
    \end{equation}
    Even when the $a_{\mum} \lsim E_{\rm keV}$ rule-of-thumb does not hold, the differential Mie scattering cross section follows roughly the same shape of the RG approximation within one $\charsig$, or~$\lsim 50-100''$ in the cases presented here (SD98).  
    The majority of light scatters within $2 \charsig$, so every observation angle $\alpha$ has a scattering angle corresponding to the furthest distance that one can see along that sight line, thereby minimizing $x$.  This condition, $\theta_{\rm sca} = \alpha /x \leq 2 \charsig$, yields the maximal time delay:
    \begin{equation}
    \label{eq:DeltaTime}
    	\delta t_{\rm max} \approx \frac{\alpha^2 D_R}{c} \left( 1 - \frac{\alpha}{2\charsig} \right) .
    \end{equation}
    Taking $\alpha \approx \charsig$ for the characteristic scattering halo image size, the total lifetime of a 1~keV halo is on the order of $10^4 (D_R/{\rm Gpc})$~years.

    \subsection{Constraints on AGN flux}
    \label{sec:AGNfluxes}

    Figure~\ref{fig:MieScatteringHalos} can be used to estimate how bright a distant AGN must be to illuminate the intergalactic medium above typical background levels.  For most observations, the {\sl Chandra} ACIS background is $\sim 0.1$~ct~s$^{-1}$ per chip, or $\mathcal{SB}_{\rm bkg} \approx 4 \times 10^{-7}$~cts~s$^{-1}$~arcsec$^{-2}$ (ACIS instrument, 0.5-2~keV).$^1$
    Far enough away from the PSF core, ghost halos should be sufficiently bright at observation angles $\alpha \approx 5''$ with $d\Psi / d\Omega \sim 10^{-5}$~arcsec$^{-2}$ for IGM dust and $\sim 10^{-8}$~arcsec$^{-2}$ for CGM dust. I calculated the threshold flux necessary to illuminate the IGM such that the scattering halo in the region of $5''$ is brighter than the background,
    \begin{equation}
    \label{eq:SBlimit}
    	C_{ps} \ \frac{d\Psi}{d\Omega}_{5''} \gsim \mathcal{SB}_{\rm bkg}
    \end{equation}
    and capable of a $3 \sigma$ detection with 50 ks of exposure time ($T_{\rm exp}$),
    \begin{equation}
    \label{eq:SNlimit}
    	\pi (5'')^2 \ \frac{d\Psi}{d\Omega}_{5''} \ \mathcal{C}_{ps} \ T_{\rm exp} \gsim 9~{\rm counts}
    \end{equation}
    where $\mathcal{C}_{ps}$ is the point source count rate in the range of 1~keV.  The resulting count rate and approximate flux requirements are listed in Table~1.

    \begin{deluxetable}{lcccc}
    \tablecolumns{5}
    \tablewidth{0pt}
    \tablecaption{AGN flux thresholds for creating observable scattering halos}
    \label{tab:HaloThresholds}
    \tablehead{
    	&
    	\multicolumn{2}{c}{\bf Above $\mathcal{SB_{\rm bkg}}$} &
    	\multicolumn{2}{c}{\bf 3$\sigma$ detection}
    	\\
    	\colhead{Dust type} &
    	\colhead{ct/s} & \colhead{flux\tablenotemark{a}} &
    	\colhead{ct/s} & \colhead{flux\tablenotemark{a}} 
    	}
	\startdata
    	Grey IGM & 0.04 & $\mathbf{3 \times 10^{-13}}$ & 0.2 & $\sim 10^{-12}$ \\
    	MRN CGM & 40 & $3 \times 10^{-10}$ & 200 & $\sim 10^{-9}$
    \enddata
    \tablenotetext{a}{1~keV photons were used to estimate the flux, which has units of erg~s$^{-1}$~cm$^{-2}$.}
    \end{deluxetable}

    Since the halo lifetime of $10^4$~years is short compared to cosmological time scales, properties of the quasar population today should describe the quasar population capable of producing X-ray scattering echoes.  All objects in the ROSAT All Sky Survey Bright Sources Catalog \citep[RASS BSC,][]{rassbsc}, which has a limiting flux around $5 \times 10^{-13}$~erg~s$^{-1}$~cm$^{-2}$, meet the minimum requirement bolded in Table~1.  
    Cross-referencing RASS BSC with the Million Quasars Catalog \citep{milliquas}, I found that approximately 5\% of these bright quasars (248 total) are far enough away ($z>1$) to expect observable IGM scattering halos.  
    I chose $z>1$ because the optical depth of the Universe rises steeply between $z = 0$ and $1$, but is relatively flat for $z > 1$ (CP12).  
    I also examined objects from the {\sl rassdssagn} catalog \citep{rassdssagn}, which cross references ROSAT sources with SDSS quasars from DR5, providing depth rather than breadth.  Approximately 15\% of the $z > 1$ AGN in {\sl rassdssagn} (156 total) are bright enough to detect IGM dust.      
    This implies that, overall, there should be $\sim 10^3$ quasars in the entire sky capable of producing observable scattering echoes from uniformly distributed, diffuse IGM dust.

    In all the catalogs examined, no sources are sufficiently bright to detect CGM dust models evaluated thus far.   I will therefore omit the CGM from further discussion in this work.

\begin{figure}[tp]
    	\centering
    	\includegraphics[scale=0.62, trim=5 15 0 20]{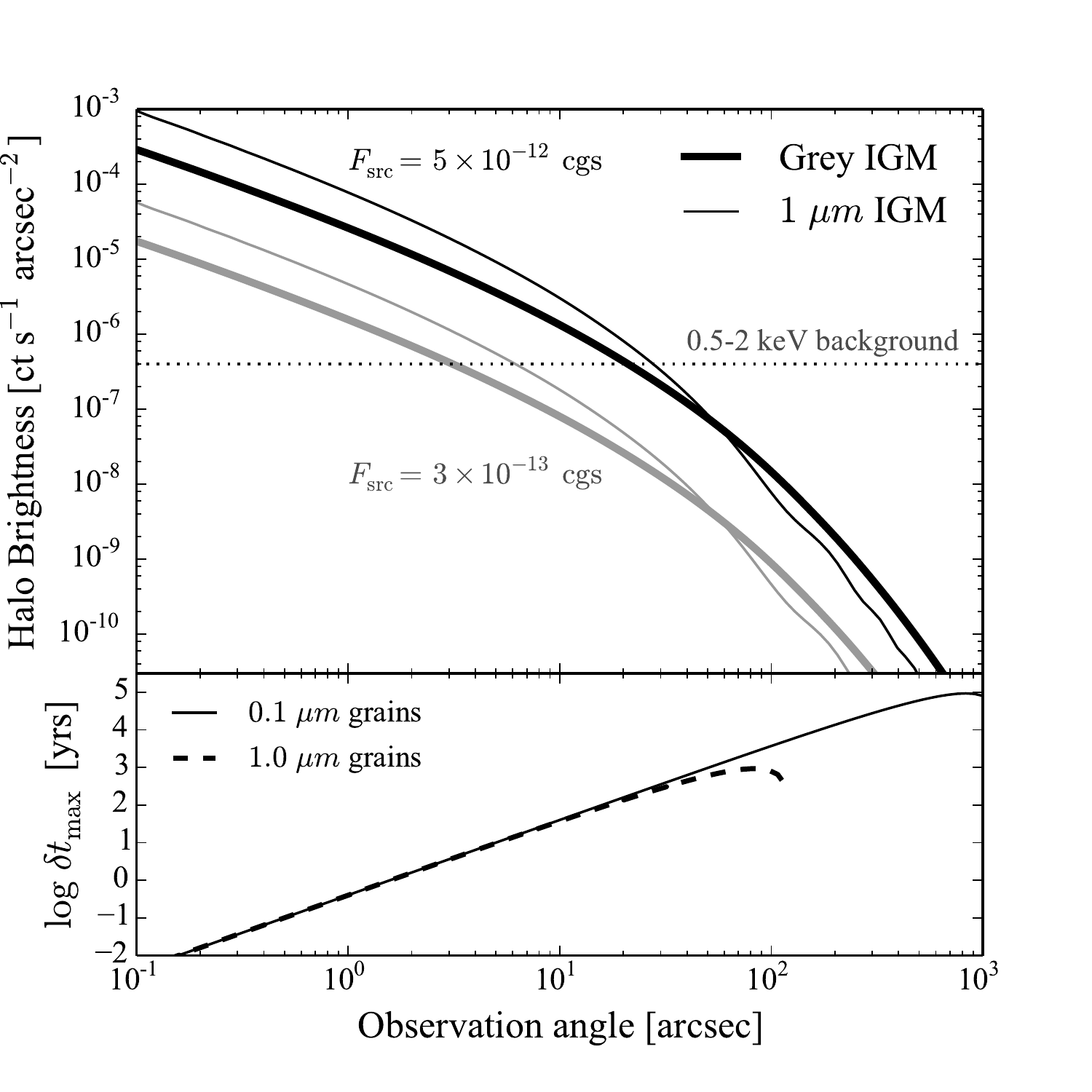}
    	\caption{ {\sl Top}:  Point source subtracted scattering halos from a uniform IGM, using the same cosmological parameters as Figure~\ref{fig:MieScatteringHalos}.  The halo brightness is directly proportional to the average AGN flux (labeled $F_{\rm src}$ in the plot).  Scattering halos from two flux thresholds ({\sl black and grey curves}) are plotted in comparison to the typical {\sl Chandra} detector background ({\sl dotted line}).
    	{\sl Bottom}:  The maximal time delay as a function of observation angle, using two limiting grain sizes.  Scattering echoes visible by {\sl Chandra} will last $\sim 1-100$~years.
    	}
    	\label{fig:HaloEchoes}
    \end{figure}

    Figure~\ref{fig:HaloEchoes} compares the {\sl Chandra} ACIS background to the predicted IGM scattering halo intensity using the minimum threshold from Table~1 and a value ten times brighter than the minimum RASS BSC flux: $5 \times 10^{-12}$~erg~s$^{-1}$~cm$^{-2}$.  There are seven $z>1$ RASS BSC quasars that meet this higher threshold.  The bottom portion of Figure~\ref{fig:HaloEchoes} shows the maximum time delay as a function of observation angle and grain size.  The observable scattering echo, which is confined to $\alpha \leq 5-20''$, should last $1 - 100$~years.

    \subsection{Constraints on AGN variability}
    \label{sec:AGNvariability}

    Hour to day-long activity, such as an X-ray afterglow to a gamma ray burst, produces an echo that probes only a tiny fraction of the total dust column.  Therefore, burst activity will not make a particularly good probe of the diffuse IGM or intervening galaxies \citep{ME1999}.
    To produce observable scattering echoes, the central X-ray point source needs to undergo a period of sustained brightness before dimming by a factor $> 1000$.

    Simulations of AGN accretion and feedback show that the supermassive black holes (SMBHs) powering quasars rarely accumulate mass at a steady rate.  
%
They may spend the majority of their time in a sub-Eddington phase, with near-Eddington accretion events triggered by the sudden infall of new gas \citep{GB2013}.  In simulations, periods of sustained activity are punctuated by major feedback events that cause the AGN to dip in luminosity by factors of $10^2 - 10^6$\citep{Hopkins2005, Novak2011}.  
%
%

    Motivated by the simulations, rapid dimming after a prolonged period of AGN activity is most likely to occur via some feedback mechanism, though technically any burst of activity longer than $10^3$~years will do.  I will call $\nu_{\rm fb}$ the characteristic frequency for such a cataclysmic event to occur.  A rough estimate for the number of scattering echoes in the sky is therefore
    \begin{equation}
    \label{eq:Nechoes}
    	N_{\rm ech} \sim 
    	\delta t_{\rm max}\ \nu_{\rm fb}\ N_{q}(F \geq F_{\rm th}, z>1)
    \end{equation}
    where $N_{q}$ is the total number of high-$z$ quasars with an apparent flux brighter than the threshold flux $F_{\rm th}$.  The above equation requires that the characteristic periodicity of the AGN growth and feedback process does not occur on a shorter timescale than the echo, i.e. $\delta t_{\rm max} < \nu_{\rm fb}^{-1}$.  Otherwise, the point source image will not remain suppressed relative to the scattering image.  This work also contains the assumption that AGN brighten slowly so that they remain at a low luminosity for a time longer than $\delta t_{\rm max}$, while feedback events are rapid (Section~\ref{sec:AccretionPhysics}).


    \begin{figure}[tp]
    	\centering
    	\includegraphics[scale=0.62, trim=5 0 0 20]{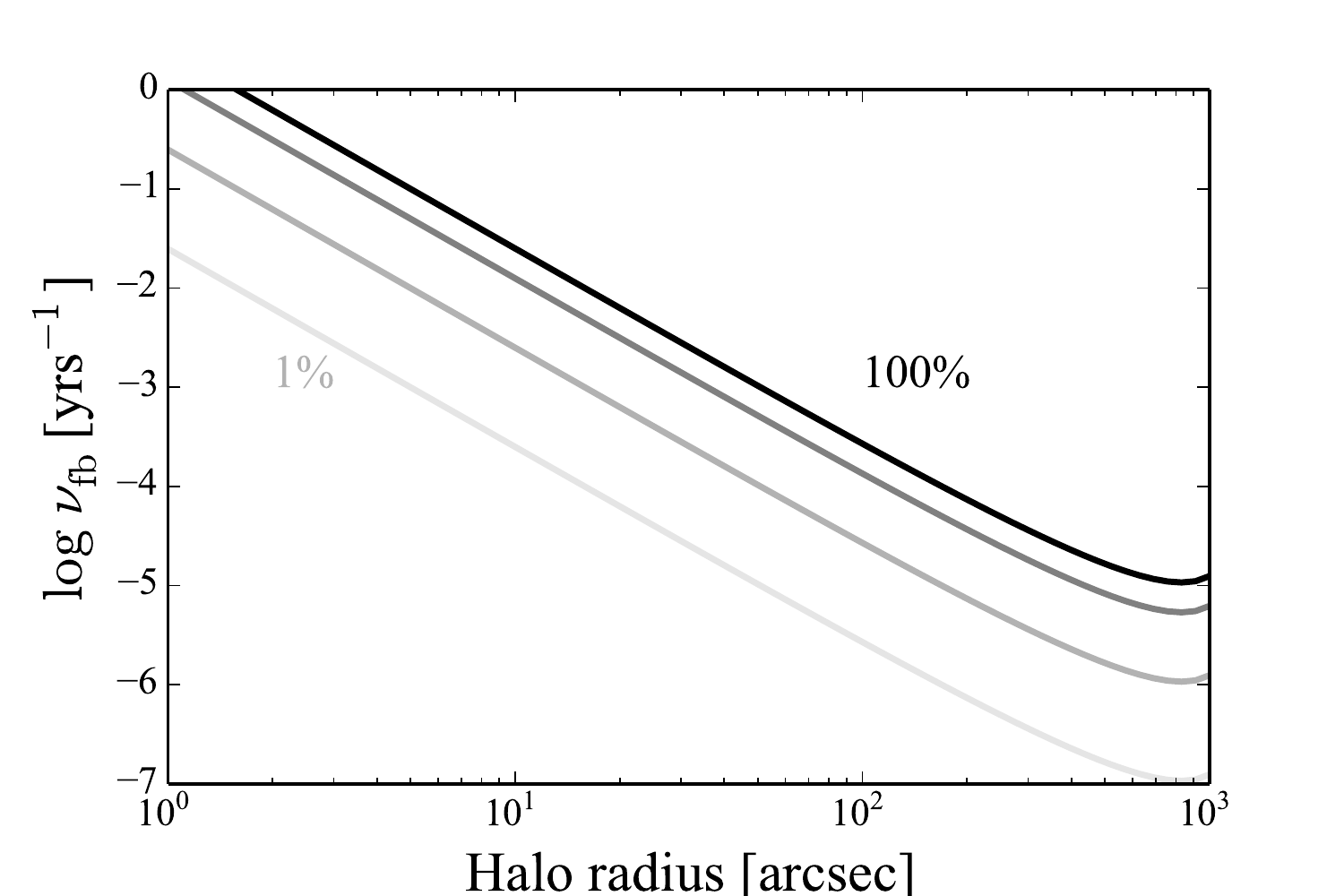}
    	\caption{ The fraction of AGN ($N_{\rm ech}/N_q$) with X-ray scattering echoes of a given size, for a particular $\nu_{\rm fb}$. The plotted contours, from black to lightest grey, are 100\%, 50\%, 10\%, and 1\%.  Beyond the 100\% contour, X-ray scattering echoes will appear as nested rings.}
    	\label{fig:fig04}
    \end{figure}
 
The number of quasars able to illuminate the IGM, $N_q (F \geq F_{\rm th}$) is strongly dependent on a number of situational factors, including the {\sl instrument} background and image exposure time.  In addition, $\delta t_{\rm max}$ and $N_q$ are linked because the brighter quasars will produce the most visibly extended halos.  For the sake of theory, I examine the abundance of scattering echoes one might expect for a perfect instrument with zero background. 
Since the inner part of the halo will vanish first, a scattering echo will qualitatively appear as a ring of surface brightness with an angular radius corresponding to the time delay $\delta t_{\rm max}$ after the feedback event.  Figure~\ref{fig:fig04} plots contours of $N_{\rm ech}/N_{\rm q}$ using $\nu_{\rm fb}$ and the halo radius calculated from $\delta t_{\rm max}$ as seen in the 0.1~$\mum$ curve in the lower portion of Figure~\ref{fig:HaloEchoes}.  Given a characteristic feedback timescale, the contours show the fraction of AGN that will have ring-like echoes of a particular radius.  
For example, if $\nu_{\rm fb} = 10^{-4}$~yrs$^{-1}$, then 10\% of AGN will have echoes $50''$ in radius, and 50\% of them will have larger but dimmer echoes about $100''$ in radius.
The $100\%$ curve in Figure~\ref{fig:fig04} marks the point at which $\delta t_{\rm max} = \nu_{\rm fb}^{-1}$.  Larger $\nu_{\rm fb}$ values will create a series of nested X-ray echo rings, analogous to those seen around variable objects \citep[e.g.][]{Tiengo2010}.


    From the catalog searches described in Section~\ref{sec:AGNfluxes},
    $$ N_{q} (F \geq 3 \times 10^{-13}, z>1) \sim 10^3 $$
and
    $$ N_{q} (F \geq 5 \times 10^{-12}, z>1) \sim 7 .$$
    With $\delta t_{\rm max}$ values of 1~year and 100~years for the two flux thresholds, respectively, I take $N_{q}~\delta t_{\rm max}$ to be roughly constant at $10^3$~years.
To achieve $N_{\rm ech} \gsim 1$, the characteristic feedback frequency must be at least one major event per 1000 years, or  $\nu_{\rm fb} \gsim 3 \times 10^{-11}$~Hz.

The characteristic frequency of interest to this work skirts the limit of human timescales.  Long-term X-ray observational campaigns covering periods up to several years ($\nu > 10^{-9}$~Hz) reveal variations on the order of $10 - 500\%$ \citep{McHardy2013, Soldi2014, Lanzuisi2014}.  In these studies, AGN flicker, but they rarely burn out.  Except for short-lived flares, 3 dex changes in luminosity are not typically observed.

There are, however, theoretical explanations to observed AGN phenomena that give hints to larger amplitude variation.
\citet{Hickox2014} produce a probability distribution over time for $L_{\rm AGN}$ between $10^{-6}$ and one~$L_{\rm Edd}$ that can explain the relatively weak correlation between AGN luminosity and the star formation rate of the host galaxy as a function of redshift.  Intermittent AGN jet activity, occurring every $10^3-10^5$~years, may also explain extended radio galaxy morphologies and the appearance of young radio sources embedded in older radio relics \citep{RB1997, Owsianik1998}.  Similar outbursts can be triggered by ionization instabilities in the accretion disk, which produce 3-4 dex luminosity swings on $10^6$~year timescales \citep{Siemiginowska1996}.  Radiation pressure instabilities, which can describe the activity observed in Galactic microquasars, would lead to AGN outbursts at $10^2 - 10^5$~year frequencies, depending on the black hole mass and accretion rate \citep{Siemiginowska2010}.  
The above results imply a range of $\nu_{\rm fb}$ values that yield $N_{\rm ech} < 10$.  

\subsection{Constraints from accretion disk physics}
\label{sec:AccretionPhysics}

The discovery of an X-ray scattering echo is constrained by yet another piece of physics: how quickly can an AGN turn off?
For an X-ray scattering echo to be visible, the source of X-rays must dim on a timescale that is shorter than $\delta t_{\rm max}$ ($\lsim 100$~years).  

Under the classic thin disk accretion model \citep{alphadisk}, the viscous timescale $t_{\rm visc} \sim R^2 / \nu_{\rm visc}$ describes how quickly material at radius $R$ will fall onto the central object, i.e. how long it will take the accretion disk to clear out.  The sound speed $c_s$ and scale height of the disk $h$ provide upper limits to the viscosity, $\nu_{\rm visc} \sim \alpha c_s h$, with $\alpha \leq 1$ being a catch-all parameter describing the contributions of magnetic fields and turbulence to the transport of angular momentum in the disk.
Using some values typical of supermassive black holes,
\begin{equation}
\label{eq:Tvisc}
	t_{\rm visc} \approx \frac{2.5\ {\rm years}}{\alpha}\ 
		R_s^{1/2}\ M_6\ T_6^{-1}
		%
		%
\end{equation}
where $M_6 = M/10^6\ {\rm M}_{\odot}$, $T_6 = T/10^6$~K, and $R_s$ is the radius in units of the Schwarszchild radius.  
For reference, the radius associated with the dynamical timescale on the order of $\delta t_{\rm max}$ is $10^5$ Schwarzschild radii for a $10^6~{\rm M}_{\odot}$ black hole, yielding $t_{\rm visc} \approx 800$~years.

To satisfy the relation $t_{\rm visc} < \delta t_{\rm max}$, a significant portion of the disk must be removed, leaving only material within $R_s \lsim 40\ M_6^{-1} (\delta t_{\rm max} / 100~{\rm yrs})$.  This inequality also implies that black holes larger than $10^7$~M$_\odot$ will not fade quickly enough to produce visible X-ray scattering echoes, at least under the alpha prescription for thin accretion disks.
%
%
It is more likely that disk instabilities, which propagate on timescales much smaller than the steady-state viscous timescale, will trigger a change fast enough to leave behind a visible X-ray scattering echo.  
The discovery of a ghost halo from the IGM would thereby place interesting limits on accretion disk physics and the nature of feedback events responsible for turning an AGN ``off.''

There is additional observational evidence that AGN can switch from a high- to low-luminosity state in a short period of time.  
A UV light echo in the vicinity of galaxy IC 2497 (``Hanny's Voorwerp'') contains fossil evidence of a nearby quasar that has dimmed by over 2-4 orders of magnitude within the last 70,000 years \citep{Lintott2009, Schawinski2010}.  
Observations of Fe K fluorescence near the Galactic Center, likely arising from a flare up of Sgr A*, indicate that our galaxy's central supermassive black hole has faded by 3-6 dex in the last 100~years \citep{Ponti2010, Capelli2012, GandoRyu2012}.  
Galactic microquasars, powered by binary accretion, can change between soft and hard states within a matter of hours \citep[e.g.][]{Greiner1996, Grinberg2013}.  While these changes do not involve factors of 1000 in luminosity, they are indicative of the variability achievable by accreting black holes.  
If these timescales are roughly proportional to black hole mass, a $10^6 - 10^7~\Msun$ AGN might change states in 10 - 100~years.

\section{Conclusion}
\label{sec:Conclusion}

Implementing Mie scattering for a Universe suffused with $\OmegaIGM \approx 3 \times 10^{-6}$ in grey dust, I have shown that only $\sim 1\%$ of X-ray light from a high-$z$ source will scatter, not $10-30\%$ as the RG-Drude approximation suggests.  
I also found that X-ray scattering off of SMC-like dust in the CGM of $z \sim 0.5$ foreground galaxies is much less likely to be observed in comparison to a dusty IGM.  
That's because uniformly distributed dust produces more centrally concentrated scattering halos.  However, the surface brightness from IGM dust scattering would still be dimmer than the {\sl Chandra} PSF wings, making it very difficult to disentangle an intergalactic scattering halo from light scattered off micro-roughness in the {\sl Chandra} mirrors.  

A natural way to get around this problem is to search for a scattering echo from an AGN that has dimmed significantly, by a factor of 1000 or more.  In this case, the point source image will be suppressed relative to the brightness of the scattering echo, which is proportional to the time-averaged flux of the AGN over $\delta t_{\rm max}$ (Equation~\ref{eq:DeltaTime}).   An AGN that has effectively turned ``off'' will not appear as a point source, and a bare scattering echo, or ghost halo, will be left behind.  Ghost halos will have observable lifetimes $\sim 1 - 100$ years, relative to the typical soft X-ray background on {\sl Chandra}.

A catalog search of X-ray bright AGN found hundreds of sources sufficiently luminous to produce $3-10''$ ghost halos that are brighter than the {\sl Chandra} ACIS background.  
These results imply a total number of scattering echoes $N_{\rm ech} \sim 10^3 (\nu_{\rm fb}/{\rm yr}^{-1})$ over the entire sky, where the frequency of major feedback events $\nu_{\rm fb}$ is by far the greatest source of uncertainty.

To leave behind a scattering echo observable by {\sl Chandra}, the contributing feedback event must meet several criteria: (i) prolonged accretion for longer than $\delta t_{\rm max}$ that produces an average apparent flux $\geq 3 \times 10^{-13}$~erg~s$^{-1}$~cm$^{-2}$, (ii) a sudden drop in luminosity by a factor $\gsim 1000$, and (iii) the time it takes the AGN to dim must be shorter than $\delta t_{\rm max}$.  The second criterion, dimming by a factor of 1000, is also necessary for an unambiguous confirmation or ruling out of the hypothesized dust density $\OmegaIGM = 3 \times 10^6$.

Simulations of AGN accretion and feedback indicate that conditions (i) and (ii) can be satisfied easily, but a rough count of eligible events yields $\nu_{\rm fb} \sim 10^{-9} - 10^{-7}$~yr$^{-1}$ \citep{Hopkins2005, Novak2011, GB2013}.  This would imply $N_{\rm ech} << 1$.  On the other hand, intermittent jet activity invoked to explain some observational properties of AGN host galaxies and radio lobes could satisfy conditions (ii) and (iii).  Several disk instability mechanisms proposed thus far imply $\nu_{\rm fb} \sim 10^{-6} - 10^{-2}$~yr$^{-1}$ \citep{RB1997, Siemiginowska1996, Siemiginowska2010}.  This produces some hope for $N_{\rm ech} \sim 1$.

The events forming a ghost scattering halo may be incredibly rare, but the discovery of such an object would have interesting implications for the fields of AGN variability, feedback, and galaxy evolution.  One of two conclusions may be reached if $N_{\rm ech} > 1$: first, that $\OmegaIGM$ is larger than previous expected, thereby increasing $N_{q}(F > F_{th})$; or second, that the frequency of AGN flickering between high- and low-luminosity accretion states is much larger than expected ($\nu_{\rm fb} \gsim 10^{-2}$~yr$^{-1}$).  The latter conclusion would also support theories of accretion beyond the thin alpha-disk prescription and help constrain the type of disk instabilities responsible for jets and other flaring activity.

The smoking gun for a ghost halo -- a diffuse, hollowed-out ring of soft X-ray light about $10''$ across -- could be imaged with {\sl Chandra}, which has sub-arcsecond imaging resolution.  The XMM EPIC detectors have a slightly lower background level than {\sl Chandra} ACIS \citep{XMMbkg} but a $6''$ FWHM, which might only barely resolve a ring structure.  
All other observatories with comparable sensitivity to soft X-rays, e.g.~Suzaku or ASCA, lack the spatial resolution necessary to identify a ring.  A ghost halo imaged with these instruments might resemble a very soft point source, with a spectrum that falls swiftly with increasing energy, i.e. a photon index $\Gamma > 2$.

Future generations of X-ray observatories are likely to achieve background rates about 5-10 times lower in comparison to {\sl Chandra} \citep{Lotti2014}, decreasing the flux threshold and thereby improving the likelihood of finding an X-ray scattering echo.  
One of the primary science goals of the {\sl Athena+} mission is to detect AGN in the $z > 6$ Universe, in order to understand formation mechanisms of the earliest supermassive black holes \citep{Athena}.  Massive $10^9~\Msun$ SMBHs discovered thus far at $z >6$ indicate that the earliest black hole seeds must have either grown at an extremely rapid pace or collapsed spontaneously.  Either way, the mechanisms responsible for forming the most massive high redshift SMBHs are not similar to the accretion we see from nearby AGN of moderate masss \citep[e.g.][]{Tanaka2009}.  Along with the instrumental increase in flux sensitivity and $5''$ imaging resolution, the increased IGM column of the high-$z$ Universe might allow X-ray scattering echoes to contribute to the {\sl Athena+} science goals by probing the variable nature of young SMBHs.
Finally, the proposed Smart-X mission would provide 30 times the effective area of {\sl Chandra} while maintaining $0.5''$ resolution \citep{SmartX}, making it the most ideal laboratory for probing X-ray scattering echoes. \\

Thank you to Frits Paerels for offering encouragement to pursue this topic.  Thank you also to Guangtun Ben Zhu, Greg Novak, and Fred Baganoff for useful conversation.  Thank you to the anonymous referee for their insightful comments, and the editors of The Astrophysical Journal for allowing a double-blind peer review process.  This work was supported in part by NASA Headquarters under the NASA Earth and Space Science Fellowship Program, grant NNX11AO09H.  The code used to calculate the scattering halo intensity is publicly available at github.com/eblur/dust; the calculations used to produce the figures in this paper are also stored at github.com/eblur/ghost\_halos in the form of iPython notebooks.

\bibliography{ghost_echoes}

\end{document}